%% file: firefighting-cs.tex
\documentclass[a4paper,UKenglish,svgnames]{lipics-v2016}
 \newcommand{\mc}{\mathcal}

\usepackage{microtype}

\title{Saving Critical Nodes with Firefighters is FPT\footnote{This work was partially supported by someone.}}
\titlerunning{Saving Critical Nodes with Firefighters is FPT} 

\author[1]{Jayesh Choudhari}
\author[1]{Anirban Dasgupta}
\author[1]{Neeldhara Misra}
\author[2]{\\M. S. Ramanujan}
\affil[1]{IIT Gandhinagar, Gandhinagar, India\\
  \texttt{(choudhari.jayesh|anirbandg|neeldhara.m)@iitgn.ac.in}}
\affil[2]{TU Wien, Vienna, Austria\\
  \texttt{ramanujan@ac.tuwien.ac.at}}
\authorrunning{J. Choudhari, A. Dasgupta, N. Misra, M.S. Ramanujan} 

\Copyright{Jayesh Choudhari, Anirban Dasgupta, Neeldhara Misra and M. S. Ramanujan}


\EventEditors{John Q. Open and Joan R. Acces}
\EventNoEds{2}
\EventLongTitle{42nd Conference on Very Important Topics (CVIT 2016)}
\EventShortTitle{CVIT 2016}
\EventAcronym{CVIT}
\EventYear{2016}
\EventDate{December 24--27, 2016}
\EventLocation{Little Whinging, United Kingdom}
\EventLogo{}
\SeriesVolume{42}
\ArticleNo{23}

\input{preamble}

\begin{document}

\maketitle

\begin{abstract}
\input{abstract.tex}
\end{abstract}

\section{Introduction}
\label{sec:introduction}
\input{ffcs-intro.tex}

\section{Preliminaries}
\label{sec:prelims}
\input{ffcs-prelims}

\section{The Parameterized Complexity of Saving a Critical Set}

\input{ffcs-section3-overview.tex}

\subsection{The FPT Algorithm}
\label{sec:FPTAlg}
\input{ffcs-fpt.tex}

\subsection{A Faster Algorithm For Trees}
\label{sec:Trees}
\input{ffcs-trees.tex}

\subsection{No Polynomial Kernel, Even on Trees}
\label{sec:npk}
\input{ffcs-npk.tex}

\section{The Spreading Model}
\label{sec:gen}
\input{ffcs-generalized.tex}

\section{Summary and Conclusions}

\input{ffcs-conclusions.tex}


\bibliography{firefighting-cs}





\end{document}

%% file: preamble.tex
\usepackage[usenames,svgnames,dvipsnames]{xcolor}
\usepackage{graphicx}
\usepackage{tikz}
\usetikzlibrary{decorations.shapes}

\usepackage{wrapfig}
\usepackage{xspace}
\usepackage[ruled,vlined,linesnumbered]{algorithm2e}
\usepackage{multicol}

\usepackage{framed}
\usepackage{boxedminipage}

\usepackage{amsmath,amssymb,amsthm,algorithm2e}
\usepackage[retainorgcmds]{IEEEtrantools}
\usepackage{pdfpages}
\usepackage{lastpage}

\usepackage{hyperref}

\hypersetup{
    pdfauthor={Neeldhara Misra},     
    pdfnewwindow=true,      
    colorlinks=true,       
    linkcolor=SlateBlue,          
    citecolor=FireBrick,        
    filecolor=magenta,      
    urlcolor=Emerald           
}

\usepackage{url}

\usepackage[backgroundcolor=Moccasin,colorinlistoftodos,disable]{todonotes}

\makeatletter
 \providecommand\@dotsep{5}
 \def\listtodoname{}
 \def\listoftodos{\@starttoc{tdo}\listtodoname}
 \makeatother

\newcounter{nmcomment}
\newcommand{\nmtodo}[1]{%
    \refstepcounter{nmcomment}%
    {%
        \todo[color={LightSteelBlue},size=\small,inline]{%
            \textbf{Comment [NM\thenmcomment ]:}~#1}%
}}

\usepackage{eulervm,euler,charter}

\newcommand{\FPT}{\textrm{\textup{FPT}}\xspace}

\newcommand{\NPH}{\textrm{\textup{NP-hard}}\xspace}
\newcommand{\NPC}{\textrm{\textup{NP-complete}}\xspace}

\newcommand{\YES}{\textsc{Yes}}
\newcommand{\NO}{\textsc{No}}

\newcommand{\problembox}[4]{
\begin{framed}
{\sc #1} \hfill Parameter: {#3} \\
\begin{tabular}{p{.15\textwidth} p{.8\textwidth}}
\hfill Input: & {#2}\\
\hfill Question: & {#4}\\
\end{tabular}
\end{framed}
}

\newcommand{\EE}{{\mathcal E}}

\newcommand{\HH}{{\mathcal H}}
\newcommand{\II}{{\mathcal I}}

\newcommand{\LL}{{\mathcal L}}

\newcommand{\OO}{{\mathcal O}}

\newcommand{\SSS}{{\mathcal S}}
\newcommand{\TT}{{\mathcal T}}

\newcommand{\WW}{{\mathcal W}}

\newcommand{\sacs}{\textsc{Saving A Critical Set}}
\newcommand{\SACS}{\textsc{SACS}}
\newcommand{\sacsg}{\textsc{Saving A Critical Set With Restrictions}}
\newcommand{\SACSG}{\textsc{SACS-R}}

%% file: abstract.tex
We consider the problem of firefighting to save a critical subset of nodes. The firefighting game is a turn-based game played on a graph, where the fire spreads to vertices in a breadth-first manner from a source, and firefighters can be placed on yet unburnt vertices on alternate rounds to block the fire. In this work, we consider the problem of saving a critical subset of nodes from catching fire, given a total budget on the number of firefighters.  

We show that the problem is para-NP-hard when parameterized by the size of the critical set. We also show that it is fixed-parameter tractable on general graphs when parameterized by the number of firefighters. 
We also demonstrate improved running times on trees and establish that the problem is unlikely to admit a polynomial kernelization (even when restricted to trees). Our work is the first to exploit the connection between
the firefighting problem and the notions of important separators and tight seperator sequences. 

Finally, we consider the spreading model of the firefighting game, a closely related problem, and show that the problem of saving a critical set parameterized by the number of firefighters is W[2]-hard, which contrasts our FPT result for the non-spreading model.

%% file: ffcs-intro.tex
The problem of Firefighting~\cite{hartnell1995firefighter} formalizes the question of designing inoculation strategies against 
a contagion that is spreading through a given network. The goal is to come up with a strategy for placing firefighters on nodes
in order to intercept the spread of fire. 
More precisely, firefighting can be thought of as a turn-based game between two players, 
traditionally the fire and the firefighter, played on a graph $G$ with a source vertex $s$. The game proceeds as follows. 

\begin{itemize}
	\item At time step $0$, fire breaks out at the vertex $s$. A vertex on fire is said to be \textit{burned}.
	\item At every odd time step $i \in \{1,3,5,\ldots\}$, when it is the turn of the firefighter, a firefighter is placed at a vertex $v$ that is not already on fire. Such a vertex is permanently \textit{protected}.
	\item At every even time step $j \in \{2,4,6,\ldots\}$, the fire spreads in the natural way: every vertex adjacent to a vertex on fire is burned (unless it was protected).
\end{itemize} 

The game stops when the fire cannot spread any more. 
A vertex is said to be \textit{saved} if there is a protected vertex on every path from $s$ to $v$. 
The natural algorithmic question associated with this game is to find a strategy that optimizes some desirable criteria, 
for instance, maximizing the number of saved vertices~\cite{cai2008firefighting}, minimizing the number of rounds, 
the number of firefighters per round~\cite{chalermsook2010resource}, or the number of burned vertices~\cite{finbow2000minimizing,cai2008firefighting}, and so on. 
These questions are well-studied in the literature, and while most variants are \NPH{}, approximation and parameterized 
algorithms have been proposed for various scenarios. See the excellent survey~\cite{Finbow:2009vma} 
as well as references within for more details. 

\nmtodo{Add citations above...}

In this work, we consider the question of finding a strategy that saves a designated subset of vertices, 
which we shall refer to as the \textit{critical set}. We refer to this problem as \sacs{} (\SACS{}) 
(we refer the reader to Section~\ref{sec:prelims} for the formal definitions). 
This is a natural objective in situations where the goal is to save specific locations as opposed to saving 
some number of them. This version of the problem has been studied by~\cite{chalermsook2010resource,king:2010cubicfirefighter,Chopin:2013tc} and is
known to be \NPH{} even when restricted to trees. 

Our aim of designing firefighting solutions in order to save a critical set is well-motivated. 
In the context of studying networked systems for instance, it is often desirable to protect
a specific set of critical infrastructure against any vulnerabilities that are cascading through the network (see~\cite{frank1970analysis}
and~\cite{ellison1997survivable} for a overview of {\em survivable network analysis} which aim to design networked systems that
survive in the face of failures by providing critical services). 
Similarly, in the context of handling widely different risk factors that a contagion might have for different sections
of the population (e.g. risk-factors that the epidemic of avian flu have for different subpopulations~\cite{cdcflu}), it is natural to ask for 
inoculation strategies to protect the identified at-risk groups.


\nmtodo{Add citations above, also a good time to talk about applications?}

\textbf{Our Contributions and Methodology.} We initiate the study of \sacs{} from a parameterized perspective. 
We first show that the problem is para-\NPH{} when parameterized by the size of the critical set, by showing that \sacs{} is 
\NPC{} even on instances where the size of the critical set is one. It is already clear from known results that \sacs{} is para-\NPH{} also when parameterized by treewidth. A third natural parameter is the number of firefighters deployed to save the critical set. Our main result is that \sacs{} is \FPT{} when parameterized by the number of firefighters, although it is not likely to have a polynomial kernel.  

Our \FPT{} algorithm is a recursive algorithm that uses the structure of tight separator sequences.  
The notion of tight separator sequences was introduced in~\cite{LokshtanovR12} and has several applications~\cite{GanianRS16,LokshtanovRSdfvs16,LokshtanovRS16} 
(some of which invoke modified definitions). A tight separator sequence is, informally speaking, a sequence of 
minimal separators such that the reachability set of $S_i$ is contained in the reachability set of $S_{i+1}$. 
Note that any firefighting solution is a $s-C$ separator, where $s$ is the source of the fire, and $C$ is the critical subset of vertices. We also obtain faster algorithms on trees by using important seperators.

As is common with such approaches, we do not directly solve \SACS{}, but an appropriately generalized form, which encodes 
information about the behavior of some solution on the ``border'' vertices, which in this case is the union of all the separators 
in the tight separator sequence.

\textbf{Related Work.} 

The Firefighting problem has received much attention in recent years. 
It has been studied in the parameterized complexity setting~\cite{cai2008firefighting,Chopin:2013tc,Cygan:2012cq,Bazgan:2014ga}
but mostly by using the number of vertices burnt or saved as parameters. 
King et.al.~\cite{king:2010cubicfirefighter} showed that for a tree of degree at most $3$, it is NP-hard
to save a critical set with budget of one firefighter per round, but is polynomial time when the fire starts from a vertex of degree at most $2$. 
Chopin~\cite{Chopin:2013tc} extended the hardness result of~\cite{king:2010cubicfirefighter} to a per-round budget $b\ge 1$
and to trees with maximum degree $b+2$.
Chalermsook et.al.\cite{chalermsook2010resource} gave an approximation to the number of firefighters per round when trying to protect a critical set. 

Anshelevich et.al.~\cite{anshelevich:2009cuts-over-time} initiated the study of the   
the spreading model, where the vaccination also spreads through the network. In Section~\ref{sec:gen} we study this problem
in the parameterized setting. 

%
%
%

\nmtodo{Refer to PC papers, critical set paper, cuts over time.}

%% file: ffcs-prelims.tex
In this section, we introduce the notation and the terminology that we will need to describe our algorithms. Most of our notation is standard. We use $[k]$ to denote the set $\{1,2,\ldots,k\}$, and we use $[k]_\OO$ and $[k]_\EE$, respectively, to denote the odd and even numbers in the set $[k]$.

\subparagraph*{Graphs, Important Separators and Tight Separator Sequences.} We introduce here the most relevant definitions, and use standard notation pertaining to graph theory based on~\cite{PCBook-CyganEtAl,Diestel:2012vm}. All our graphs will be simple and undirected unless mentioned otherwise. For a graph $G = (V,E)$ and a vertex $v$, we use $N(v)$ and $N[v]$ to refer to the open and closed neighborhoods of $v$, respectively. The \emph{distance} between vertices $u,v$ of $G$ is the length of a shortest path from $u$ to $v$ in $G$; if no such path exists, the distance is defined to be $\infty$. A graph $G$ is said to be \emph{connected} if there is a path in $G$ from every vertex of $G$ to every other vertex of $G$. If $U\subseteq V$ and $G\left[U\right]$ is connected, then $U$ itself is said to be connected in $G$.  For a subset $S \subseteq V$, we use the notation $G\setminus S$ to refer to the graph induced by the vertex set $V \setminus S$. 

The following definitions about important separators and tight separator sequences will be relevant to our main \FPT{} algorithm. We first define the notion of the reachability set of a subset $X$ with respect to a subset $S$. 

\begin{definition}[Reachable Sets]
	Let $G = (V, E)$ be an undirected graph, let $X \subseteq V$ and $S \subseteq V \setminus X$. We denote by $R_G(X,S)$ the set of vertices of $G$ \textit{reachable} from $X$ in $G \setminus S$ and by $NR_G(X, S)$ the set of vertices of $G$ not reachable from $X$ in $G \setminus S$. We drop the subscript $G$ if it is clear from the context.
\end{definition}

We now turn to the notion of an $X$-$Y$ separator and what it means for one separator to cover another. 

\begin{definition}[Covering by Separators]
	Let $G = (V, E)$ be an undirected graph and let $X, Y \subset V$ be two disjoint vertex sets. A subset $S \subseteq V \setminus (X \cup Y )$ is called an $X-Y$ separator in $G$ if $R_G(X, S) \cap Y = \emptyset$, or in other words, there is no path from $X$ to $Y$ in the graph $G \setminus S$. We denote by $\lambda_G (X, Y)$ the size of the smallest $X-Y$ separator in $G$. An $X-Y$ separator $S_1$ is said to \textit{cover} an $X-Y$ separator $S$ with respect to $X$ if $R(X,S_1) \supset R(X,S)$. If the set $X$ is clear from the context, we just say that $S_1$ covers $S$. An $X-Y$ separator is said to be inclusionwise minimal if none of its proper subsets is an $X-Y$ separator.
\end{definition}

If $X = \{x\}$ is a singleton, then we abuse notation and refer to a $x-Y$ separator rather than a $\{x\}-Y$ separator. A separator $S_1$ dominates $S$ if it covers $S$ and is not larger than $S$ in size:

\begin{definition}[Dominating Separators \cite{CyganFKLMPPS15}]
	Let $G = (V, E)$ be an undirected graph and let $X, Y \subset V$ be two disjoint vertex sets. An $X-Y$ separator $S_1$ is said to \textit{dominate} an $X-Y$ separator $S$ with respect to $X$ if $|S_1| \leq |S|$ and $S_1$ covers $S$ with respect to $X$. If the set $X$ is clear from the context, we just say that $S_1$ dominates $S$.
\end{definition}

We finally arrive at the notion of important separators, which are those that are not dominated by any other separator.

\begin{definition}[Important Separators \cite{CyganFKLMPPS15}] Let $G = (V, E)$ be an undirected graph, $X, Y \subset V$ be disjoint vertex sets and $S \subseteq V \setminus (X \cup Y)$ be an $X-Y$ separator in $G$. We say that $S$ is an \textit{important} $X-Y$ separator if it is inclusionwise minimal and there does not exist another $X-Y$ separator $S_1$ such that $S_1$ dominates $S$ with respect to $X$. \end{definition}

%

It is  useful to know that the number of important separators is bounded as an \FPT{} function of the size of the important separators. 

\begin{lemma} \label{lem:compute_imp_sep}\cite{CyganFKLMPPS15} Let $G = (V, E)$ be an undirected graph, $X, Y \subset V$ be disjoint vertex sets of $G$. For every $k \geq 0$ there are at most $4^k$ important $X-Y$ separators of size at most $k$. Furthermore, there is an algorithm that runs in time $O(4^kk(m + n))$ which enumerates all such important X-Y separators, where $n = |V|$ and $m = |E|$.
\end{lemma}

We are now ready to 
%
%
%
recall the notion 
of tight separator sequences introduced in~\cite{LokshtanovR12}. However, the definition and structural 
lemmas regarding tight separator sequences used in this paper are closer to that from~\cite{LokshtanovRS16}. 
Since there are minor modifications in the definition as compared to the one in \cite{LokshtanovRS16}, we  give the requisite proofs for the sake of completeness.

\begin{definition}\label{def:smallest separator sequence} 
Let $X$ and $Y$ be two subsets of $V(G)$ and let $k\in {\mathbb N}$. A \emph{tight ($X$,$Y$)-reachability sequence} of order $k$ is 
an ordered collection ${\cal H} = \{H_0,H_1,\dots,H_q\}$ of sets in $V(G)$ satisfying the following properties: 
\begin{itemize}
\item $X \subseteq H_i \subseteq V(G) \setminus N[Y]$ for any $0 \leq i\leq q$; 
\item $X = H_0 \subset H_1\subset H_2\subset \dots \subset H_q$;
\item for every $0\leq i\leq q$, $H_i$ is reachable from $X$ in $G[H_i]$  and every vertex in $N(H_i)$ can reach $Y$ in $G - H_i$ \\
(implying that $N(H_i)$ is a minimal ($X$,$Y$)-separator in $G$);
\item $|N(H_i)| \leq k$ for every $1 \leq i \leq q$;
\item $N(H_i) \cap N(H_j) = \emptyset$ for all $1 \leq i,j \leq q$ and $i \neq j$;
\item For any $0\leq i\leq q-1$, there is no ($X$,$Y$)-separator $S$ of size at most $k$ 
where $S\subseteq H_{i+1}\setminus N[H_{i}]$ or $S \cap N[H_q]= \emptyset$ or $S \subseteq H_1$.
\end{itemize} 
We let  $S_i = N(H_i)$, for $1 \leq i \leq q$, $S_{q + 1} = Y$, and $\mc{S} = \{S_0, S_1, \dots, S_q, S_{q+1}\}$. 
We call $\mc{S}$ a \emph{tight ($X$,$Y$)-separator sequence} of order $k$. 
\end{definition} 

\begin{lemma}(see for example  \cite{LokshtanovRS16})\label{lem:compute-sep-seq}\label{lem:find_tight_sequence}
There is an algorithm that, given an $n$-vertex $m$-edge graph $G$, subsets $X, Y\in V(G)$ and an integer $k$, runs in time $O(k mn^2)$ and either 
correctly concludes that there is no ($X$,$Y$)-separator of size at most $k$ in $G$ or returns the sets 
$H_0,H_1,H_2\setminus H_1,\dots, H_q\setminus H_{q-1}$ corresponding to a tight ($X$,$Y$)-reachability sequence ${\cal H}=\{H_0,H_1,\dots,H_q\}$ of order $k$.
\end{lemma}

\begin{proof}
The algorithm begins by checking whether there is an $X$-$Y$ separator of size at most $k$. If there is no such separator, then it simply outputs the same. Otherwise, it uses the algorithm of Lemma \ref{lem:compute_imp_sep} to compute an arbitrary important $X$-$Y$ separator $S$ of size at most $k$ such that there is no $X$-$Y$ separator of size at most $k$ that covers $S$.

Although the algorithm of Lemma \ref{lem:compute_imp_sep} requires time $O(4^kk(m + n))$ to enumerate \emph{all} important $X$-$Y$ separators of size at most $k$, \emph{one}  important separator of the  kind described in the previous paragraph can in fact be computed in time $O(kmn)$ by the same algorithm. 

If there is no $X$-$S$ separator of size at most $k$, we stop and return the set $R(X,S)$ as the only set in a tight $(X,Y)$-reachability sequence. 
Otherwise, we recursively compute a tight $(X,S)$-reachability sequence ${\cal P}=\{P_0,\dots, P_r\}$ of order $k$ and define ${\cal Q}=\{P_0,\dots, P_r,R(X,S)\}$ as a tight $(X,Y)$-reachability sequence of order $k$. It is straightforward to see that all the properties required of a tight $(X,Y)$-reachability sequence are satisfied. Finally, since the time required in each step of the recursion is $O(kmn)$ and the number of recursions is bounded by $n$, the number of vertices, the claimed running time follows.
\end{proof}

\subparagraph{Saving a Critical Set.} We now turn to the definition of the firefighting problem. The game proceeds as described earlier: we are given a graph $G$ with a vertex $s \in V(G)$. To begin with, the fire breaks out at $s$ and vertex $s$ is burning. At each step $t \geq 1$, first the firefighter protects one vertex not yet on fire - this vertex remains permanently protected - and the fire then spreads from burning vertices to all unprotected neighbors of these vertices. The process stops when the fire cannot spread anymore. In the definitions that follow, we formally define the notion of a firefighting strategy. 

\begin{definition}\label{def:ff-strategy}[Firefighting Strategy] A $k$-step firefighting strategy is defined as a function $\mathfrak{h}: [k] \rightarrow V(G)$. 	Such a strategy is said to be \textit{valid in $G$ with respect to $s$} if, for all $i \in [k]$, when the fire breaks out in $s$ and firefighters are placed according to $\mathfrak{h}$ for all time steps up to $2(i-1)-1$, the vertex $\mathfrak{h}(i)$ is not burning at time step $2i-1$, and the fire cannot spread anymore after timestep $2k$. If $G$ and $s$ are clear from the context, we simply say that $\mathfrak{h}$ is a valid strategy.
\end{definition}

\begin{definition}[Saving $C$] For a vertex $s$ and a subset $C \subseteq V(G) \setminus \{s\}$, a firefighting strategy $\mathfrak{h}$ is said to save $C$ if $\mathfrak{h}$ is a valid strategy and $\cup_{i=1}^k \mathfrak{h}(i)$ is a $\{s\}$-$C$ separator in $G$, in other words, there is no path from $s$ to any vertex in $C$ if firefighters are placed according to $\mathfrak{h}$.	
\end{definition}

We are now ready to define the parameterized problem that is the focus of this work.

\problembox{\sacs{}~(\SACS{})}{An undirected $n$-vertex graph $G$, a vertex $s$, a subset $C \subseteq V(G) \setminus \{s\}$, and an integer k.}{$k$}{Is there a valid $k$-step strategy that saves $C$ when a fire breaks out at $s$?}

\nmtodo{Need to define the spreading model and the $b$-firefighters-per-step model here as well, assuming we go in that direction.}

\subparagraph*{Parameterized Complexity.} We follow standard terminology pertaining to parameterized algorithms based on the monograph~\cite{PCBook-CyganEtAl}. Here we define a known technique to prove kernel lower bounds, called cross composition. Towards this, we first define polynomial equivalence relations. 

\begin{definition}[polynomial equivalence relation~\cite{BJK11}] \label{polyEquivalenceRelation}
An equivalence relation~${\cal R}$ on~$\Sigma^*$, where $\Sigma$ is a finite alphabet, is called a \emph{polynomial equivalence relation} if the following holds:
$(1)$ equivalence of any~$x,y \in \Sigma^*$ can be checked in time polynomial in~$|x|+|y|$, 
 and $(2)$	any finite set~$S \subseteq \Sigma^*$ has at most~$(\max _{x \in S} |x|)^{O(1)}$~equivalence classes.
\end{definition}

\begin{definition}[cross-composition~\cite{BJK11}] \label{crossComposition}
Let~$L \subseteq \Sigma^*$ and let~$Q \subseteq \Sigma^* \times \mathbb{N}$ be a parameterized problem. We say that~$L$ \emph{cross-composes} into~$Q$ if there is a polynomial equivalence relation~${\cal R}$ and an algorithm which, given~$t$ strings~$x_1, x_2, \ldots, x_t$ belonging to the same equivalence class of~${\cal R}$, computes an instance~$(x^*,k^*) \in \Sigma^* \times \mathbb{N}$ in time polynomial in~$\sum _{i=1}^t |x_i|$ such that: (i) $(x^*, k^*) \in Q \Leftrightarrow x_i \in L$ for some $1 \leq i \leq t$  and (ii) $k^*$ is bounded by a polynomial in ($\max _{1\leq i\leq t} |x_i|+\log t$).
\end{definition}

The following theorem allows us to rule out the existence of a polynomial kernel for a parameterized problem.

\begin{theorem}[\cite{BJK11}]
\label{theorem:conp:crosscomposition}
If an \NPH problem~$L\subseteq \Sigma^*$ has a cross-composition into the parameterized problem~$Q$ and~$Q$ has a polynomial kernel then NP $\subseteq$ coNP/poly.
\end{theorem}


%% file: ffcs-section3-overview.tex
In this section, we describe the \FPT{} algorithm for \sacs{} and our cross-composition construction for trees. The starting point for our \FPT{} algorithm is the fact that every solution to an instance $(G,s,C,k)$ of \SACS{} is in fact a $s$-$C$ separator of size at most $k$. Although the number of such separators may be exponential in the size of the graph, it is a well-known fact that the number of \textit{important} separators is bounded by $4^k n^{O(1)}$~\cite{}. For several problems, one is able to prove that there exists a solution that is in fact an important separator. In such a situation, an \FPT{} algorithm is immediate by guessing the important separator. 

In the \SACS{} problem, unfortunately, there are instances where none of the solutions are important separators. However, this approach turns out to be feasible if we restrict our attention to trees, leading to improved running times. This is described in greater detail in Section~\ref{sec:Trees}. Further, in Section~\ref{sec:npk}, we also show that we do not expect \SACS{} to admit a polyonimal kernel under standard complexity-theoretic assumptions. We establish this by a cross-composition from \SACS{} itself, using the standard binary tree approach, similar to~\cite{Bazgan:2014ga}.

We describe our \FPT{} algorithm for general graphs in Section~\ref{sec:FPTAlg}. This is an elegant recursive procedure that operates over tight separator sequences, exploiting the fact that a solution can never be contained entiely in the region ``between two consecutive separators''. Although the natural choice of measure is the solution size, it turns out that the solution size by itself cannot be guaranteed to drop in the recursive instances that we generate. Therefore, we need to define an appropriate generalized instance, and work with a more delicate measure. We now turn to a detailed description of our approach.

We note that the \SACS{} problem is para-\NPC{} when parameterized by the size of the critical set, by showing that the problem is already \NPC{} when the critical set has only one vertex.

\begin{theorem}
	\SACS{} is \NPC{} even when the critical set has one vertex. 
\end{theorem}

\begin{proof}
Let $(G, k)$ be an instance of $k-CLIQUE$ problem. We construct a graph $G'$ as follows. For each edge $(u, v) \in E(G)$ create a node $s_{uv}$. We denote this set of nodes by $E$. For each node $v \in V(G)$ create a node $v \in G'$ and denote this set of nodes by $V$. Add two nodes $s$ and $t$, where $s$ is the node on which the fire starts and $t \in C$ (CRITICAL SET), the node to be saved. Connect $t$ to all the nodes in set $E$. Connect $s$ to each node $v \in V$ by a path of length $k$. Let us refer these nodes, which are on the paths from $s$ to $V$, as $V_{-1}$. Create $(m - {k \choose 2} - 1)$ copies of set $V$. Denote these copies as $V_1, V_2, ... , V_{m-{k \choose 2} -1}$. Let $V_m = \cup_{i=1}^{{m-{k \choose 2} -1}}V_i$ and $V$ be denoted by $V_0$. Add an edge $(v_{i,j}, v_{i,j+1})$ for $v_i \in V_j$ and $v_i \in V_{j+1}$ for all $0 \leq j < (m-{k \choose 2}-1)$. For each edge $e=(u,v)$ in $E(G)$ add an edge $(u, s_{u,v})$ and $(v, s_{u,v})$ where $u, v \in V_{m-{k \choose 2}-1}$ and $s_{u,v} \in E$. Now set $k' = k + m - {k \choose 2}$.
\end{proof}

\begin{lemma}
\label{atmost-k}
	At most $k$ firefighters can be placed on the nodes in set $(V_{-1} \cup V_0)$ in $G'$.
\end{lemma}
\begin{proof}
	As per the construction of $G'$, each node $v \in V_0$ is connected to $s$ with a path of length $k$ and each node $u \in V_{-1}$ is at a distance of length less than $k$. Thus, there are at most $k$ time steps at which firefighters can be employed on the nodes in $V_{-1} \cup V_0$.

\begin{lemma}
\label{not-less-than-k}
	Protecting less than $k$ nodes from the set $H_k = (V_{-1} \cup V_{0} \cup V_{m})$ in $G'$ is not a successful strategy to save $t$.
\end{lemma}
\begin{proof}
	To save $t$, all the nodes in $E$ ($|E| = m$) must either be protected or saved. Suppose we protect $\gamma < k$ nodes in set $H_k$. Observe that, protecting $\gamma$ nodes in $H_k$ can save at most $\gamma \choose 2$ nodes in $E$. Therefore, in order to save $t$ we need to protect/save remaining $m - {\gamma \choose 2}$ nodes in $E$. Given the constraint on the budget as well as the number of time instances we are left with, we can protect $m - {k \choose 2} + (k - \gamma)$ more nodes to save $t$. If $\alpha = m - {k \choose 2} + (k - \gamma)$ and $\beta = m - {\gamma \choose 2}$ then, the claim is $\beta - \alpha > 0$.

	$$ \beta - \alpha = m - {\gamma \choose 2} - \bigg(m - {k \choose 2} + (k - \gamma)\bigg)$$

	$$ = {k \choose 2} - {\gamma \choose 2} + (\gamma - k) = \frac{k (k - 1)}{2} - \frac{\gamma (\gamma - 1)}{2} + (\gamma - k)$$

	$$ = \frac{k^2 - k - \gamma^2 + \gamma}{2} + (\gamma - k)$$
	$$ = \frac{(k+\gamma)(k - \gamma) - (k - \gamma)}{2} + (\gamma - k)$$
	$$ = (k-\gamma) \bigg[\frac{k + l}{2} - \frac{1}{2} - 1\bigg] > 0$$

	Therefore, it is not possible to save $t$ if we protect $\gamma < k$ nodes in $H_k$, as this requires $m - {\gamma \choose 2}$ more nodes to be protected, which is not feasible.

\end{proof}

	We claim that SAVING A CRITICAL SET with $|C|=1$ on $(G', k', s, C={t})$ is a \YES{}-instance if and only if $k-CLIQUE$ is an \YES{}-instance in $(G, k)$.

	Suppose $G$ has a $k-clique$ denoted as $K$. Then the firefighting strategy is to protect the vertices $v_i \in V_0$ corresponding to the vertices $v_i \in K$. Protecting these $k$ vertices guarantees to save ${k \choose 2}$ vertices in the set $E$. Also, from Lemma \ref{atmost-k} it follows that protecting these $k$ vertices is valid with respect to time constraints. The remaining $m - {k \choose 2}$ in $E$ can be protected at each allowed time step after placing the $k$ firefighters in $V_0$. This as well is valid with respect to the time constraints as the set of nodes $E$ are at a distance $m - {k \choose 2}$ from the set of nodes $V_0$.  

	Suppose that $G'$ has a valid firefighting strategy $S = \{u_1, u_2, \dots, u_{k+m-{k \choose 2}}\} $ with $k + m - {k \choose 2}$ firefighters. If a firefighter is placed on any node $u \in V_{-1}$ or on any node $w \in V_m$, then it is equivalent to placing a firefighter on the node $v_i \in V_0$ to which the nodes $u$ and $w$ have a path. Therefore, in the firefighting strategy $S$ if there is a firefighter on node $u \in V_{-1}$ then it is pushed to $v_i \in V_0$ such that $u-v_i$ is a path, and if there is a firefighter on node $w \in V_m$, then it is pulled back to $v_j \in V_0$ such that $w-v_j$ is a path. Also, it follows from Lemma \ref{not-less-than-k} that for a successful firefighting strategy, $k$ firefighters must be placed in $H_k = (V_{-1} \cup V_{0} \cup V_{m})$. Therefore, these $k$ nodes in $V_0$ on which the firefighters are placed (by pushing from $u \in V_{-1}$ or pulling from $w \in V_m$), must form a clique in $G$ with the $k \choose 2$ edges corresponding to the nodes saved in the set $E$. If these $k$ nodes do not form a clique, then with the remaining $m-{k\choose 2}$ nodes it won't be possible to save/cover all the vertices in $E$.	
	
\end{proof}

%% file: ffcs-fpt.tex
Towards the \FPT{} algorithm for \SACS{}, we first define a generalized firefighting problem as follows. In this problem, in addition to $(G,s,C,k)$, we are also given the following:

\begin{itemize}
	\item $P \uplus Q \subseteq [2k]_\OO$, a set of \textit{available time steps},
	\item $Y \subset V(G)$, a subset of \textit{predetermined firefighter locations}, and
	\item a bijection $\gamma: Q \rightarrow Y$, a \textit{partial strategy} for $Q$.
\end{itemize}

The goal here is to find a valid partial $k$-step firefighting strategy over $(P \cup Q)$ that is consistent with $\gamma$ on $Q$ and saves $C$ when the fire breaks out at $s$. We assume that no firefighters are placed during the time steps $[2k]_\OO \setminus (P \cup Q)$. For completeness, we formally define the notion of a valid partial firefighting strategy over a set.

\begin{definition}[Partial Firefighting Strategy] A \textit{partial} $k$-step firefighting strategy on $X \subseteq [2k]_{\OO}$ is defined as a function $\mathfrak{h}: X \rightarrow V(G)$. 	Such a strategy is said to be \textit{valid in $G$ with respect to $s$} if, for all $i \in X$, when the fire breaks out in $s$ and firefighters are placed according to $\mathfrak{h}$ for all time steps upto $[i-1]_{\OO} \cap X$, the vertex $\mathfrak{h}(i)$ is not burning at time step $i$. If $G$ and $s$ are clear from the context, we simply say that $\mathfrak{h}$ is a valid strategy over $X$. 
\end{definition}

What it means for partial strategy to save $C$ is also analogous to what it means for a strategy to save $C$. The only difference here is that we save $C$ despite not placing any firefighters during the time steps $j$ for $j \in [2k]_{\OO}\setminus X$.

\begin{definition}[Saving $C$ with a Partial Strategy] For a vertex $s$ and a subset $C \subseteq V(G) \setminus \{s\}$, a partial firefighting strategy $\mathfrak{h}$ over $X$ is said to save $C$ if $\mathfrak{h}$ is a valid strategy and $\cup_{i \in X} \mathfrak{h}(i)$ is a $s-C$ separator in $G$, in other words, there is no path involving only burning vertices from $s$ to any vertex in $C$ if the fire starts at $s$ and firefighters are placed according to $\mathfrak{h}$.	
\end{definition} 

The intuition for considering this generalized problem is the following: when we recurse, we break the instance $G$ into two parts, say subgraphs $G^\prime$ and $H$. An optimal strategy for $G$ employs some firefighters in $H$ at some time steps $X$, and the remaining firefighters in $G^\prime$ at time steps $[2k]_\OO \setminus X$. When we recurse, we would therefore like to achieve two things:

\begin{itemize}
	\item Capture the interactions between $G^\prime$ and $H$ when we recursively solve $H$, so that a partial solution that we obtain from the recursion aligns with the larger graph, and
	\item Constrain the solution for the instance $H$ to only use time steps in $X$, ``allowing'' firefighters to work in $G^\prime$ for the remaining time steps. 
\end{itemize}

The constrained time steps in our generalized problem cater to the second objective, and the predetermined firefighter locations partially cater to the first. We now formally define the generalized problem. 

\problembox{\sacsg{}~(\SACSG{})}{An undirected $n$-vertex graph $G$, vertices $s$ and $g$, a subset $C \subseteq V(G) \setminus \{s\}$, a subset $P \uplus Q \subseteq [2k]_\OO$, $Y \subset V(G)$ (such that $|Y| = |Q|$, $2k-1 \in Q$ and $g \in Y$), a bijeciton $\gamma: Q \rightarrow Y$ such that $\gamma(2k-1) = g$, and an integer $k$.}{$k$}{Is there a valid partial $k$-step strategy over $P \cup Q$ that is consistent with $\gamma$ on $Q$ and that saves $C$ when a fire breaks out at $s$?}

We use $p$ and $q$ to denote $|P|$ and $|Q|$, respectively. Note that we can 
solve an \SACS{} instance $(G,s,C,k)$ by adding an isolated vertex $g$ and solving the \SACSG{} 
instance $(G,s,C,2k+2,g,P,Q,Y,\gamma)$, where $P = [2k]_{\OO}$, $Q = \{2k+1\}$, $Y = \{g\}$ and $\gamma(2k+1) = g$. 
Therefore, it suffices to describe an algorithm that solves \SACSG{}. 
The role of the vertex $g$ is mostly technical, and will be clear in due course. 

We now describe our algorithm for solving an instance $\II := (G,s,C,k,g,P,Q,Y,\gamma)$ of \SACSG{}. Throughout this discussion, for the convenience of analysis of \YES{} instances, let $\mathfrak{h}$ be an arbitrary but fixed valid partial firefighting strategy in $G$ over $P \cup Q$, consistent with $\gamma$ on $Q$, that saves $C$. Our algorithm is recursive and works with pieces of the graph based on a tight $s-C$-separator sequence of separators of size at most $|P|$ in $G \setminus Y$. We describe the algorithm in three parts: the pre-processing phase, the generation of the recursive instances, and the merging of the recursively obtained solutions.

\subparagraph*{Phase 0 --- Preprocessing.} Observe that we have the following easy base cases:

\begin{itemize}
	\item If $G \setminus Y$ has no $s-C$ separators of size at most $p$, then the algorithm returns \NO{}.
	\item If $p = 0$, then we have a \YES{}-instance if, and only if, $s$ is separated from $C$ in $G \setminus Y$ and $\mathfrak{h} := \gamma$ is a valid partial firefighting strategy over $Q$. In this case, the algorithm outputs \YES{} or \NO{} as appropriate. 
	\item If $p > 0$ and $s$ is already separated from $C$ in $G \setminus Y$, then we return \YES{}, since any arbitrary partial strategy over $P \cup Q$ that is consistent with $\gamma$ on $Q$ is a witness solution. 
\end{itemize}

If we have a non-trivial instance, then our algorithm proceeds as follows. To begin with, we compute a tight $s-C$ separator sequence of order $p$ in $G \setminus Y$. Recalling the notation of Definition~\ref{def:smallest separator sequence} , we use $S_0, \ldots, S_{q+1}$ to denote the separators in this sequence, with $S_0$ being the set $\{s\}$ and $S_{q+1} = C$. We also use $W_0, W_1, \ldots, W_q, W_{q+1}$ to denote the reachability regions between consecutive separators. More precisely, if $\HH$ is the tight $s-C$ reachability sequence associated with $\SSS$, then we have: 

\[W_i := H_i\setminus N[H_{i-1}] \text{ for } 1 \leq i \leq q,\] 


while $W_{q+1}$ is defined as $G \setminus (N[H_q] \cup C)$. We will also frequently employ the following notation: 

$$\SSS  = \bigcup_{i=1}^q S_i \mbox{ and } \WW  = \bigcup_{i=1}^{q+1} W_i.$$

This is a slight abuse of notation since $\SSS$ is also used to denote the sequence $S_0, \ldots, S_{q+1}$, but the meaning of $\SSS$ will typically be clear from the context. 

We first observe that if $q > k$, the separator $S_q$ can be used to define a valid partial firefighting strategy. The intuition for this is the following: since every vertex in $S_q$ is at a distance of at least $k$ from $s$, we may place firefighters on vertices in $S_q$ in any order during the available time steps. Since $|S_q| \leq p$ and $S_q$ is a $s-C$ separator, this is a valid solution. Thus, we have shown the following:

\begin{lemma}\label{lem:short-seq}
	If $G$ admits a tight $s-C$ separator sequence of order $q$ in $G \setminus Y$ where $q > k$, then $\II$ is a \YES{}-instance.
\end{lemma}

Therefore, we return \YES{} if $q > k$ and assume that $q \leq k$ whenever the algorithm proceeds to the next phase. 

This concludes the pre-processing stage. 

\nmtodo{Should the above be converted into an explicit proposition?}

\subparagraph*{Phase 1 --- Recursion.} Our first step here is to guess a partition of the set of available time steps, $P$, into $2q+1$ parts, denoted by $A_0, \ldots, A_q, A_{q+1}$ and $B_1, \ldots, B_{q+1}$. The partition of the time steps represents how a solution might distribute the timings of its firefighting strategy among the sets in $\SSS$ and $\WW$.  The set $A_i$ denotes our guess of $\cup_{v \in S_i}\mathfrak{h}^{-1}(v)$ and $B_j$ denotes our guess of $\cup_{v \in W_j}\mathfrak{h}^{-1}(v)$. Note that the number of such partitions is $(2q+1)^p \leq (2k+1)^k$. We define $g_0(k) := (2k+1)^k$. We also use $\TT_1(P)$ to denote the partition $A_0, \ldots, A_q$  and $\TT_2(P)$ to denote $B_0, \ldots, B_{q+1}$. 

We say that the partition $(\TT_1(P),\TT_2(P))$ is non-trivial if none of the $B_i$'s are such that $B_i = P$. Our algorithm only considers non-trivial partitions --- the reason this is sufficient follows from the way tight separator sequences are designed, and this will be made more explicit in Lemma~\ref{lem:measure-drop} in due course.

Next, we would like to guess the behavior of a partial strategy over $P$ restricted to $\SSS$. Informally, we do this by associating a signature with the strategy $\mathfrak{h}$, which is is a labeling of the vertex set with labels corresponding to the status of a vertex in the firefighting game when it is played out according to $\mathfrak{h}$. Every vertex is labeled as either a vertex that had a firefighter placed on it, a burned vertex, or a saved vertex. The labels also carry information about the earliest times at which the vertices attained these statuses. More formally, we have the following definition.

\begin{definition}
Let $\mathfrak{h}$ be a valid $k$-step firefighting strategy (or a partial strategy over $X$). The signature of $\mathfrak{h}$ is defined as a labeling $\mathfrak{L}_\mathfrak{h}$ of the vertex set with labels from the set:

$$\LL = \left(\{\mathfrak{f}\} \times X \right) \cup \left( \{\mathfrak{b}\} \times [2k]_{\EE}  \right)  \cup  \{\mathfrak{p}\} ,$$

where:

\begin{equation*} \mathfrak{L}_\mathfrak{h}(v) = 
\begin{cases} 

(\mathfrak{f},t) & \text{if } \mathfrak{h}(t) = v,\\ 

(\mathfrak{b},t) & \text{if } t \text{ is the earliest time step at which } v \text{ burns}  ,\\

\mathfrak{p} & \text{if } v \text{ is not reachable from } s \text{ in } G \setminus \left( \{ \mathfrak{h}(i) ~|~ i \in [2k]_{\OO} \} \right)  \\

\end{cases} \end{equation*}

\end{definition}



We use array-style notation to refer to the components of $\mathfrak{L}(v)$, for instance, if $\mathfrak{L}(v) = (\mathfrak{b},t)$, then $\mathfrak{L}(v)[0] = \mathfrak{b}$ and $\mathfrak{L}(v)[1] = t$. The algorithm begins by guessing the restriction of $\mathfrak{L}_\mathfrak{h}$ on $\SSS$, that is, it loops over all possible labellings: 

$$\mathfrak{T}: \SSS \rightarrow \left(\{\mathfrak{f}\} \times P \right) \cup \left( \{\mathfrak{b}\} \times [2k]_\EE  \right)  \cup  \{\mathfrak{p}\}.$$ 

The labeling $\mathfrak{T}$ is called legitimate if, for any $u \neq v$,  whenever $\mathfrak{T}(u)[0] = \mathfrak{T}(v)[0] = \mathfrak{f}$, we have $\mathfrak{T}(u)[1] \neq \mathfrak{T}(v)[1]$. We say that a labeling $\mathfrak{T}$ over $\SSS$ is compatible with $\TT_1(P) = (A_0,\ldots,A_q)$ if we have:

\begin{itemize}
	\item for all $0 \leq i \leq r$, if $v \in S_i$ and $\mathfrak{h}(v)[0] = \mathfrak{f}$, then $\mathfrak{h}(v)[1] \in A_i$.
	\item for all $0 \leq i \leq r$, if $t \in A_i$, there exists a vertex $v \in S_i$ such that $\mathfrak{h}^{-1}(\mathfrak{f},t) = v$.
\end{itemize}  

The algorithm considers only legitimate labelings compatible with the current choice of $\TT_1(P)$. By Lemma~\ref{lem:short-seq}, we know that any tight $s-C$ separator sequence considered by the algorithm at this stage has at most $k$ separators of size at most $p$ each. Therefore, we have that the number of labelings considered by the algorithm is bounded by $g_1(k) := (p+k+1)^{(kp)} \leq (3k)^{O(k^2)} \leq k^{O(k^2)}$.

We are now ready to split the graph into $q+1$ recursive instances. For $1 \leq i \leq q+1$, let us define $G_i = G[S_{i-1} \cup W_i \cup S_i \cup Y]$. Also, let $\mathfrak{T}_{i} := \mathfrak{T}\mid_{V(G_i) \cap \SSS}$. Notice that when using $G_i$'s in recursion, we need to ensure that the independently obtained solutions are compatible with each other on the non-overlapping regions, and consistent on the common parts. We force consistency by carrying forward the information in the signature of $\mathfrak{h}$ using appropriate gadgets, and the compatibility among the $W_i$'s is a result of the partitioning of the time steps. 

Fix a partition of the available time steps $P$ into $\TT_1(P)$ and $\TT_2(P)$, a compatible labeling $\mathfrak{T}$ and $1 \leq i \leq q+1$. We will now define the \SACSG{} instance $\II\langle i,\TT_1(P),\TT_2(P),\mathfrak{T}_i \rangle$. Recall that $\II = (G,s,C,k,g,P,Q,Y,\gamma)$. To begin with, we have the following:

\begin{itemize}
	\item Let $X_i = A_{i-1} \cup A_i$ and let $P_i = B_i$. 
	\item Let $Q_i := X_i \cup Q$ and $Y_i := Y \cup X_i$. We define $\gamma_i$ as follows:
	
\begin{equation*} \gamma_i(t) = 
\begin{cases} 

\gamma(t) & \text{if } t \in Q,\\ 

v & \text{if } t \in X_i \text{ and } \mathfrak{T}_i(v) = (\mathfrak{f},t)
\end{cases} \end{equation*}
\end{itemize}

Note that $\gamma_i$ is well-defined because the labeling was legitimate and compatible with $\TT_1(P)$. We define $H_i$ to be the graph $\chi(G_i,\mathfrak{T}_i)$, which is described below. 

\begin{itemize}
	\item To begin with, $V(H_i) = V(G_i) \cup \{s^\star,t^\star\}$
	\item Let $v \in V(G_i)$  be such that $\mathfrak{T}_i(v)[0] = \mathfrak{b}$. Use $\ell$ to denote $\mathfrak{T}_i(v)[1]/2$. Now, we do the following:
		\begin{itemize}
			\item  Add $k+1$ internally vertex disjoint paths from $s^\star$ to $v$ of length $\ell + 1$, in other words, these paths have $\ell - 1$ internal vertices. 
			\item Add $k+1$ internally vertex disjoint paths from $v$ to $g$ of length $k - \ell - 1$.
		\end{itemize}
	\item Let $v \in V(G_i)$  be such that $\mathfrak{T}_i(v) = \mathfrak{p}$. Add an edge from $v$ to $t^\star$.
	\item We also make $k+1$ copies of the vertices $t^\star$ and all vertices that are labeled either burned or saved. This ensures that no firefighters are placed on these vertices. 
\end{itemize}

For $1 \leq i \leq q+1$, the instance $\II\langle i,\TT_1(P),\TT_2(P),\mathfrak{T}_i \rangle$ is now defined as $(\chi(G_i,\mathfrak{T}_i),s^\star,C = \{t^\star\}, k, g, P_i, Q_i, Y_i, \gamma_i)$.

\subparagraph*{Phase 2 --- Merging.} Our final output is quite straightforward to describe once we have the $\mathfrak{h}[\mathfrak{T}_i,i]$'s. Consider a fixed partition of the available time steps $P$ into $\TT_1(P)$ and $\TT_2(P)$, and a labeling $\mathfrak{T}$ of $\SSS$ compatible with $\TT_1(P)$. If all of the $(q+1)$ instances $\II\langle i,\TT_1(P),\TT_2(P),\mathfrak{T}_i \rangle$, $1 \leq i \leq q+1$ return \YES{}, then we also return \YES{}, and we return \NO{} otherwise. Indeed, in the former case, let $\mathfrak{h}[i,\TT_1(P),\TT_2(P),\mathfrak{T}]$ denote a valid partial firefighting strategy for the instance $\II\langle i,\TT_1(P),\TT_2(P),\mathfrak{T}_i \rangle$. We will show that $\mathfrak{h}^\star$, described as follows, is a valid partial firefighting strategy that saves $C$.

\begin{itemize}
\item For the time steps in $Q$, we employ firefighters according to $\gamma$. 	
\item For the time steps in $\TT_1(P)$, we employ firefighters according to $\mathfrak{T}$. This is a well-defined strategy since $\mathfrak{T}$ is a compatible labeling.
\item For all remaining time steps, i.e, those in $\TT_2(P) = \{B_1, \ldots, B_{q+1}\}$, we follow the strategy given by $\mathfrak{h}[i,\TT_1(P),\TT_2(P),\mathfrak{T}]$. 
\end{itemize}

It is easily checked that the strategy described above agrees with $\mathfrak{h}[i,\TT_1(P),\TT_2(P),\mathfrak{T}]$ for all~$i$. Also, the strategy is well-defined, since $\TT_1(P)$ and $\TT_2(P)$ form a partition of the available time steps. Next, we will demonstrate that $\mathfrak{h}^\star$ is indeed a valid strategy that saves $C$, and also analyze the running time of the algorithm.

\begin{algorithm}[H]
\SetKwInput{KwData}{Input}
 \KwData{An instance $(G,s,C,k,g,P,Q,Y,\gamma)$, $p := |P|$}
 \KwResult{\YES{} if $\II$ is a \YES{}-instance of \SACSG{}, and \NO{} otherwise.}
\lIf{$p = 0$ and $s$ and $C$ are in different components of $G \setminus Y$}{\Return{\YES{}}}
\lElse{\Return{\NO}}
\lIf{$p > 0$ and $s$ and $C$ are in different components of $G \setminus Y$}{\Return{\YES{}}}
\lIf{there is no $s-C$ separator of size at most $p$}{\Return{\NO{}}}
Compute a tight $s-C$ separator sequence $\SSS$ of order $p$.\\
\lIf{the number of separators in $\SSS$ is greater than $k$}{\Return{\YES{}}}
\Else{
\For{a non-trivial partition $\TT_1(P),\TT_2(P)$ of $P$ into $2q+1$ parts}{
\For{a labeling $\mathfrak{T}$ compatible with $\TT_1(P)$}{
\lIf{$\bigwedge_{i=1}^{q+1}$(Solve-\SACSG{}($\II \langle i,\TT_1(P),\TT_2(P),\mathfrak{T}_i \rangle$))}{\Return{\YES{}}}
}
}
\Return{\NO{}}
}
\caption{Solve-\SACSG{}($\II$)}
\end{algorithm}

\subsubsection*{Correctness of the Algorithm} 

We first show that the quantity $p$ always decreases when we recurse. 

\begin{lemma}\label{lem:measure-drop}
Let $\II^\prime := \II\langle i,\TT_1(P),\TT_2(P),\mathfrak{T}_i \rangle = (\chi(G_i,\mathfrak{T}_i),s^\star,C = \{t^\star\}, k, g, P_i, Q_i, Y_i, \gamma_i)$ be an arbitrary but fixed instance constructed by Solve-\SACSG{}($\II$). Then, $|P_i| < |P|$.
\end{lemma}

\begin{proof}
The claim follows from the fact that $P_i = B_i$, and $(\TT_1(P),\TT_2(P))$ is a non-trivial partition of $P$. 
\end{proof}

\begin{lemma}\label{lem:correct-f}
If $\II$ is a \YES{} instance of \SACSG{}, then our algorithm returns \YES{}.
\end{lemma}

\begin{proof} (Sketch.) The statement follows by induction on $p$, which allows us to assume the correctness of the output of the recursive calls.  The correctness of the base cases is easily checked. If $\II$ is a \YES{} instance admits a solution $\mathfrak{h}$, then it induces a partition $\TT_1(P), \TT_2(P)$ of $P$. We argue that this is always a non-trivial partition and is therefore considered by the algorithm. Indeed, suppose not. This would imply that $\mathfrak{h}$ places all its firefighters in $W_i$ for some $1 \leq i \leq q+1$. However, this implies that $D := \cup_{t \in P_i} \mathfrak{h}(t) \subseteq W_i$ is a $s-C$ separator in $G \setminus Y$ that is entirely contained in $W_i$, which contradicts the definition of a tight $s-C$ separator sequence. 

Now, define the labeling $\mathfrak{T}$ by projecting the signature of $\mathfrak{h}$ on $\SSS$. Clearly, this labeling is compatible with $\TT_1(P)$ (by definition). We claim that all the instances $\II\langle i,\TT_1(P),\TT_2(P),\mathfrak{T}_i \rangle$ are \YES{} instances. Indeed, it is easy to check that the projection of $\mathfrak{h}$ on the subgraph $G_i$ is a valid solution for the instance $\II\langle i,\TT_1(P),\TT_2(P),\mathfrak{T}_i \rangle$. It follows that the algorithm returns \YES{} since these $(q+1)$ instances return \YES{} by the induction hypothesis. 
\end{proof}

For the remainder of our discussion on correctness, our goal will be to prove the following reverse claim.

\begin{lemma}\label{lem:correct-b}
If the output of the algorithm is \YES{}, then $\II$ is a \YES{} instance of \SACSG{}.
\end{lemma}

This lemma is shown by establishing that $\mathfrak{h}^\star$ is indeed a valid solution for \SACS{}. We arrive at this conclusion by a sequence of simple claims. We fix the accepting path in the algorithm, that is, an appropriate partition of $P$ and the labeling $\mathfrak{T}$ on $\SSS$ that triggered \YES{} output. Our first claim says that the recursive instances respect the behavior dictated by the labeling that they were based on. 

\begin{lemma}\label{lem:burn}
Let $1 \leq i \leq q+1$ and $v \in \SSS \cap H_i$. If $\mathfrak{T}_i(v) = (\mathfrak{b},t)$, then in any partial strategy employed on the instance $\II\langle i,\TT_1(P),\TT_2(P),\mathfrak{T}_i \rangle$ that saves the critical set, the vertex $v$ burns exactly at time step $t$.
\end{lemma}

\begin{proof}
Let $\mathfrak{h}^\prime$ be an arbitrary but fixed valid partial strategy for $\II\langle i,\TT_1(P),\TT_2(P),\mathfrak{T}_i \rangle$ that saves the critical set. Assume that $v$ burns with respect to $\mathfrak{h}^\prime$. Let $t^\prime$ be the earliest time step at which $v$ burns with respect to  $\mathfrak{h}^\prime$. The graph $\chi(G_i,\mathfrak{T}_i)$ contains $(k+1)$ vertex-disjoint paths from the source to $v$ of length $t/2$, which ensures that $t^\prime \leq t$. However, if $t^\prime < t$, then the vertex $g$ catches fire at time step $2k-1$ because of the $(k+1)$ vertex-disjoint paths of length $k-t/2-1$ that are present from $v$ to $g$. This implies that $t^\star$ cannot be saved if $v$ burns earlier than $t$. The other case is that $v$ does not burn with respect to $\mathfrak{h}^\prime$. The only way for this to happen is if a firefighter is placed on $v$, however, since the instance has $k+1$ copies of $v$, we have that at least one copy of $v$ burns, and the claim follows. 
\end{proof}

Our next claim is that vertices that are labeled saved never burn with respect to $\mathfrak{h}^\star$.

\begin{lemma}\label{lem:save}
For any $v \in \SSS$, if $\mathfrak{T}(v) = \mathfrak{p}$, then $v$ does not burn with respect to $\mathfrak{h}^\star$.
\end{lemma}

\begin{proof}
We prove this by contradiction. Let $P$ be a path from $s$ to $v$ where all vertices on $P$ are burning. Let $v^\prime$ be the first vertex on this path which is such that $\mathfrak{T}(v^\prime) = \mathfrak{p}$, and let $u$ be the last vertex on this path which is before $v^\prime$ and that intersects a separator. Observe that $u$ is well-defined unless $v^\prime \in S_1$, which is a special case that we will deal with separately. Note that $u$ must be a vertex that is labeled as burned (by the choice of $v^\prime$). Therefore, the path from $u$ to $v$ is present in a recursive instance, where we know that $v^\prime$ is adjacent to a critical vertex, which contradicts the fact that we defined $\mathfrak{h}^\star$ based on valid strategies that save critical sets in the recursive instances. If $v^\prime \in S_1$ and $u$ is not well-defined, then observe that there is a direct path from $s$ to $v^\prime$ in the first recursive instance, and the same argument applies. 
\end{proof}

We finally show, over the next two claims, that the function $\mathfrak{h}^\star$ is a valid strategy that saves the critical set. 

\begin{lemma}\label{lem:valid}
	The function $\mathfrak{h}^\star$ is a valid partial strategy over $(P \cup Q)$ for the instance $\II$. 
\end{lemma}
\begin{proof} We prove this by contradiction, and we also assume that Lemma~\ref{lem:correct-b} holds for all recursive instances (by induction on $p$). If $\mathfrak{h}^\star$ is not a valid strategy, then there exists some $1 \leq i \leq q+1$ for which there is a vertex $v \in W_i$ and a time step $t \in B_i$ such that $\mathfrak{h}^\star(t) = v$ and $v$ is burning at time step $t$. Consider the path $P$ from $s$ to $v$. Let $u$ be the last vertex on the path $P$ that intersects $S_{i-1} \cup S_i$. Observe that $u$ is a vertex that is either labeled by $(\mathfrak{b},t)$ or $\mathfrak{p}$. This leads to the following two scenarios:

\begin{itemize}
	\item In the first case, the part of the path from $u$ to $v$ is present in the instance $\II\langle i,\TT_1(P),\TT_2(P),\mathfrak{T}_i \rangle$, because of the agreement of $\mathfrak{h}^\star$ and $\mathfrak{h}[i,\TT_1(P),\TT_2(P),\mathfrak{T}]$ and Lemma~\ref{lem:burn}. Therefore, $\mathfrak{h}[i,\TT_1(P),\TT_2(P),\mathfrak{T}]$ was not a valid firefighting strategy.
	\item The second situation implies that a vertex labeled protected is burning in $\II$ when $\mathfrak{h}^\star$ is employed, which contradicts Lemma~\ref{lem:save}. 
\end{itemize} 
\end{proof}

\begin{lemma}\label{lem:safe}
	The partial strategy $\mathfrak{h}^\star$ saves the set $C$ in the instance $\II$. 
\end{lemma}
\begin{proof}(Sketch.) The proof of this is similar to the proof of Lemma~\ref{lem:valid}. Any burning path $P$ from $s$ to any vertex in $C$ must intersect $S_q$. We let $v$ be the last vertex from $S_q$ on $P$, and observe that $v$ must have a $(\mathfrak{b},t)$. Therefore, the path from $v$ to the vertex in $C$ exists in $\II\langle q+1,\TT_1(P),\TT_2(P),\mathfrak{T}_i \rangle$, and also burns in the same fashion, since $\mathfrak{h}^\star$ agrees with $\mathfrak{h}[q+1,\TT_1(P),\TT_2(P),\mathfrak{T}]$. Again, this contradicts the accuracy of the algorithm on the recursive instance. 
	\end{proof}

The proof of Lemma~\ref{lem:correct-b} now follows from Lemmas~\ref{lem:valid} and~\ref{lem:burn}.

\input{rta.tex}

%% file: rta.tex
\subsubsection*{Running Time Analysis} 

We show that our algorithm runs in time $f(k)^{p^2}O(n^2m)$, where $f(k) = k^{O(k)}$.  Indeed, observe that the running time of the algorithm is governed by the following recurrence:

\[ T(n,m,k,p) \leq O(n^2mp) + (p+k+1)^{kp}\sum_{i=1}^{q+1}T(n_i,m_i,k,p_i) \]

where the term $(p+k+1)^{kp}$ denotes an upper bound on the product of the total number of non-trivial partitions of $P$ into $(2q+1)$ parts and the total number of legitimate labelings of $\SSS$ compatible with $\TT_1(P)$. The first term accounts for checking the base cases and the running time of computing a tight separator sequence of order $p$ (see Lemma~\ref{lem:compute-sep-seq}). Since $p$ is always at most $k$, we rewrite the recurrence as 

\[ T(n,m,k,p) \leq O(n^2mp) + f(k)^{p}\sum_{i=1}^{q+1}T(n_i,m_i,k,p_i), \]

  
 where $f(k)=k^{O(k)}$. Furthermore, observe that $\sum_{i=1}^{q+1} n_i \leq n + pk$. The first inequality follows from the fact that every vertex in $\SSS$ appears in at most two recursive instances. The first occurrence is counted in $n$, while all the second occurrences combined amount to at most $pk$. 

Finally, recall that $p_i < p$ by Lemma~\ref{lem:measure-drop} and  $p \leq k$ by definition (since $P \subseteq [2k]_\OO$). Therefore, the depth of the recursion is bounded by $p$, and the time spent at each level of recursion is  proportional to $k^{O(kp)}n^2m$. This implies the claimed running time. 

Based on the analysis above and Lemmas~\ref{lem:correct-f} and \ref{lem:correct-b}, we have thus demonstrated the main result of this section.

\begin{theorem}
	\SACS{} is \FPT{} and has an algorithm with running time  $f(k)O(nm^2)$, where $f(k) = k^{O(k^3)}$.
\end{theorem}

%% file: ffcs-trees.tex
In this section we consider the setting when the input graph $G$ is a tree. WLOG, we consider the
vertex $s$ to be the root of the tree.  We first state an easy claim 
that shows that WLOG, we can consider the critical set to be the leaves. The proof of the following lemma follows from the fact that the firefighting
solution has to be a $s-C$ seperator. 
\begin{lemma}
\label{lem:subtree}
When the input graph $G$ is a tree, if there exists a solution to \SACS{}, there exists a solution such that all firefighter locations
are on nodes that are on some path from $s$ to $C$. 
\end{lemma}

Given the above claim, our algorithm to construct a firefighting solution is the following-- exhaustively search all the important $s-C$ separators
that are of size $k$. For each vertex $v$ in a separator $Y$, we place firefighters on $Y$ in the increasing order of distance from $s$ and check 
whether this is a valid solution. 
The following lemma claims that if there exists a firefighting solution, the above algorithm will return one. 
\begin{lemma}
Solving the \SACS{}  problem for input graphs that are trees takes time  $O^*(4^k)$.  
\end{lemma}
\begin{proof}
 Using Lemma~\ref{lem:subtree}, it is enough to consider the subtree $T$ that contains nodes only on $s-C$ paths. The critical set $C$ 
 is then a subset of the leaves of $T$. Suppose $Y\subset V(T)$ contains the locations for a solution to the firefighter problem. 
 WLOG, $Y$ is a minimal $s-C$ separator. 
 Consider $I$ which is an important separator that dominates $Y$. 
 Clearly, $|I| \le |Y| \le k$, and $I$ is also a minimal separator. 
  
  For each $x \in I$, define $S_x$ to be the set of $y \in Y$ such that $y$ lies on the (unique) $s-x$ path. Note
  that each $y\in Y$ must belong to some $s-C$ path, and all $s-C$ paths have some node $x\in I$. Furthermore,
  $R(s, Y) \subseteq R(s, I)$, which means that each $y\in Y$ lies on some $s-I$ path. 
  It follows that $Y \subseteq \cup_{x\in I} S_x$. Finally, by the minimality of $Y$, each 
  $S_x$ satisfies $|S_x| \le 1$, since otherwise we could remove one of the nodes of $S_x$ from $Y$. 
  Hence,
  \begin{align*}
   |Y| \le |\cup_{x\in I} S_x| \le \sum_{x\in I} |S_x| \le |I| \le |Y|
  \end{align*}
  Thus  $|\cup_{x\in I} S_x| = \sum_{x\in I} |S_x|$, and  
  it follows that $S_x \cap S_u = \emptyset$ for all $x\neq u$. From $\sum_{x\in I} |S_x| = |I|$ and $|S_x| \le 1$ for each $x$
  it also follows that for each $x$, $|S_x| = 1$.
  
  We then design a firefighting solution using $I$ in the following manner. For each node $x\in I$, the firefighter in 
  location $x$ is placed whenever the original solution (using $Y$) placed a firefighter on the unique node in $S_x$. 
  Since the node in $S_x$ is closer to $s$ than $x$, the location $x$ is still available at this step. Hence this is a valid strategy. 
  Note that if there is any valid placement ordering for $I$, then the placement 
  ordering according to increasing distance from $s$ is also valid. 
 
  The claim then follows by noting that enumerating all the important separators of size at most $k$ takes time $O^*(4^k)$. 
 
\end{proof}

%% file: ffcs-npk.tex
Given that there is a FPT algorithm for \SACS{} when restricted to trees, in this section we show that \SACS{} on trees has no polynomial kernel. As mentioned before, the proof technique used here is on the similar lines of the proof showing no polynomial kernel for SAVING ALL BUT k-VERTICES by Bazgan et. al.\cite{Bazgan:2014ga}. 

\begin{theorem}
	\SACS{} when restricted to trees does not admit polynomial kernel, unless NP $\subseteq$ coNP/poly.
\end{theorem}

To prove this theorem, we will use the Definitions (\ref{polyEquivalenceRelation}, \ref{crossComposition}) mentioned in section \ref{sec:prelims}. We use Theorem \ref{theorem:conp:crosscomposition}, for which we consider \SACS{} on trees as analogous to language $L$, which is shown to be NP-complete when the critical set $C$ is the set of all leaves \cite{king:2010cubicfirefighter}. First we give a lemma that we will be using in the proof.

\begin{lemma}
\label{lemma:more-than-one-leaves-burn}
	For a given instance of \SACS{} $(T, r, C, k)$, where $T$ is a full binary tree with height $h$ and $k=h$, if more than one vertex is protected at a depth $d\leq h$, then more than one leaf burns.
\end{lemma}

\begin{proof}
	Consider a case when more than one firefighter is placed at a depth $d \leq h$. It is easy to see that at most 1 firefighter can be placed at depth 1. Therefore, more than 1 firefighter will always be at depth $d \geq 2$. Also, at any depth $d \geq 1$, there are $2^d$ nodes, which is strictly greater than the number of odd time steps at which the firefighters are allowed to be placed. This says that, all the nodes at a particular depth $d\geq 1$ cannot be protected by firefighters. Therefore, at any depth $d \geq 2$, there are at least two subtrees which are unprotected and can be burnt. And thus, given the constraint that at most one firefighter can be placed at any allowed time step, there is at least one subtree which is burnt and can never be completely protected/saved by the firefighters.

	However, one of the strategies to let only one leaf burn, is to fix a path from root to a leaf and keep protecting the siblings of the nodes on that path. This will be more clearer as we go ahead in the proof of the current theorem.
\end{proof}

\begin{lemma}
	The unparameterized version of \SACS{} restricted to trees cross composes to \SACS{} restricted to trees when parameterized by the number of firefighters.
\end{lemma}
\begin{proof}
	We take an appropriate equivalence relation ${\cal R}$ such that ${\cal R}$ puts all the malformed instances into one class and all the well formed instances are grouped into equivalence classes according to the number of vertices of the tree, and number of firefighters (parameter $k$) required to save the critical set $C$. We assume that we are given a sequence of $t$ instances $(T_i, s_i, C_i, k)_{i=1}^{t}$  of the unparameterized version of \SACS{} restricted to trees, each  rooted at $s_i$. Note that each of the $t$ instances belong to the same equivalence class i.e. for all the instances $k$ is same. Also, consider that $t = 2^h$, for some $h \geq 1$, else, we duplicate some instances and the duplication at most doubles the number of instances.

	Using these $t$ instances, we create a new full binary tree $T'$ as follows. Let $T'$ be rooted at $s'$, and $h$ be the height of $T'$~$(2^h = t)$. For each leaf $i \in \{1, \dots, t$\}, replace the $i^{th}$ leaf by the instance $(T_i, s_i, C_i, k)$ and now, set $k' = k+h = k + \log t$. Observe that, for $T'$ the set of all leaves is the union of the leaves of the instances $T_i$ i.e. $\cup_{i=1}^{t} C_i$.

	To prove the correctness, we show that the tree $T'$ formed by the composition of $t$ instances saves all the leaves with $k'$ firefighters if and only if there exits at least one instance $T_i$ for $i \in \{1,2, \dots, t\}$, that saves its critical set $C_i$ (i.e. the set of all leaves) with $k$ firefighters.


	Suppose, $T'$ has a successful firefighting strategy. Then, from lemma \ref{lemma:more-than-one-leaves-burn}, it follows that, the firefighting strategy that saves all but one root $s_i$ of $T_i$ is the one that protects exactly one vertex at each depth of the tree. This costs exactly $\log(t) = h$ firefighters. Thus, after $2h$ time steps, there is exactly one vertex $s_i$ (root of instance $T_i$) which is on fire. And the critical set $C_i$ for the instance $(T_i, s_i, C_i, k)$ is saved using $k$ firefighters. 

	Now suppose, there is an instance $(T_i, s_i, C_i, k)$ that has a successful firefighting strategy with $k$ firefighters. Thus, the leaves $C_i$ of this instance are saved by the firefighting strategy with $k$ firefighters. Now, the goal is to save all other leaves $(\cup_{j=1}^t C_j)\setminus C_i$. $T'$ being a tree, there is an unique path from the root $s$ to the node $s_i$ (i.e. the root of the instance $T_i$). Denote the path as $P = (s, v_1, v_2, \dots, v_{\log(t)-1}, s_i)$. Note that each node $v_j$ is at depth $j$ and $s_i$ is at depth $\log(t)$. Let $u_1, u_2, \dots, u_{\log(t)-1}, s_j$ be the siblings of the nodes $v_1, v_2, \dots, v_{\log(t)-1}, s_i$ on the path $P$ respectively.  Now, the firefighting strategy that protects the sibling $u_j$ of node $v_j$ at time step $2j-1$ and sibling $s_j$ of the node $s_i$ at time step $2h-1 = 2\log(t)-1$ saves all other leaves of $T'$.  
\end{proof}


%% file: ffcs-generalized.tex
	The spreading model for firefighters was defined by Anshelevich et al. \cite{anshelevich:2009cuts-over-time} as ``Spreading Vaccination Model''. In contrast to the firefighting game described in Section 1, in the spreading model, the firefighters (vaccination) also spread at even time steps as similar to that of the fire. That is, at any even time step if there is a firefighter at node $v_i$, then the firefighter extends (vaccination spreads) to all the neighbors of $v_i$ which are not already on fire or are not already protected by a firefighter. Consider a node $v_i$ which is not already protected or burning at time step $2j$. If $u_i$ and $w_i$ are neighbors of $v_i$, such that, $u_i$ was already burning at time step $2j-1$ and $w_i$ was protected at time step $2j-1$, then at time step $2j$, $v_i$ is protected. That is, in the spreading model the firefighters dominate or win over fire. For the spreading model, the firefighting game can be defined formally as follows:
	\begin{itemize}
		\item At time step $0$, fire breaks out at the vertex $s$. A vertex on fire is said to be \textit{burned}.
		\item At every odd time step $i \in \{1,3,5,\ldots\}$, when it is the turn of the firefighter, a firefighter is placed at a vertex $v$ that is not already on fire. Such a vertex is permanently \textit{protected}.
		\item At every even time step $j \in \{2,4,6,\ldots\}$, first the firefighter extends to every adjacent vertex to a vertex protected by a firefighter (unless it was already protected or burned), then the fire spreads to every vertex adjacent to a vertex on fire (unless it was already protected or burned). Needless to say, the vertices protected at even time steps are also permanently \textit{protected}.
	\end{itemize} 

	In the following theorem, we show that in spite of the spreading power that the firefighters have, \SACS{} is hard.

\begin{theorem}\label{thm:whard}
	In the spreading model, \SACS{} is as hard as $k$-\textsc{Dominating Set}.
\end{theorem}

\begin{proof}
	Let $(G, k)$ be an instance of $k$-DOMINATING SET problem. We construct a graph $G'$ as follows. Add $2$ copies $V^{(1)}$ and $V^{(2)}$ of the nodes $V(G)$ in $G'$, i.e. for each node $v_i \in V(G)$ add nodes $v_i^{(1)}$ and $v_i^{(2)}$. Add a vertex $s$, the vertex from where the fire breaks out. For each vertex $v_i^{(1)} \in V^{(1)}$, add a path of length $k$ from $s$ (i.e. a path from $s$ to $v_i^{(1)}$ would be like $(s, u_{i_1}, \dots, u_{i_{k-1}}, v_i^{(1)})$). Similarly, for each edge $(v_i, v_j) \in E(G)$, add a path of length $k$ from $v_i^{(1)}$ to $v_j^{(2)}$, and a path of length $2k$ from $v_i^{(2)}$ to  $v_j^{(1)}$ in $G'$. Also, for each $v_i \in V(G)$ add a path of length $2k$ from $v_i^{(1)}$ to $v_i^{(2)}$. Let $C = V^{(2)}$ be the critical set and $k' = k$.

	We claim that, \SACS{} on $(G', s, k', C=V^{(2)})$ is a \YES{}-instance if and only if $k$-\textsc{Dominating Set} is a \YES{} instance on $(G,k)$.

	Suppose that, $G$ has a $k$-\textsc{Dominating Set} $K$. Then the strategy that saves the critical set $C$ is the one that protects the vertices $v_i^{(1)} \in V^{(1)}$ corresponding to the vertices $v_i$ in the dominating set $K$. The nodes $v_i \in K$ being the dominating set, the corresponding nodes $v_i^{(1)} \in V^{(1)}$ dominate all the nodes $v_j^{(2)} \in V^{(2)}$. And thus, the firefighters on the nodes $v_i^{(1)} \in V^{(1)}$ corresponding to the dominating set $K$, eventually extend firefighters to all the nodes in $V^{(2)}$ before fire reaches to any node $v_j^{(2)} \in V^{(2)}$. Also, protecting the vertices $v_i^{(1)} \in V^{(1)}$ corresponding to $v_i \in K$ is a valid firefighting strategy as per the Definition \ref{def:ff-strategy}.

	Suppose that $\mathfrak{h}: [k] \rightarrow V(G')$ is an optimal  firefighting strategy for $(G', s, k', C=V^{(2)})$. Let $S = \{u_1, \dots, u_k\}$ be the set of vertices which are protected by the firefighting strategy, i.e. $\mathfrak{h}(i) = u_i$ for $i \in [k]$. Note that, as per the definition of the firefighting game, $i^{th}$ firefighter is placed at time step $2i-1$. Lets denote the paths from the nodes in $V^{(1)}$ to the nodes in $V^{(2)}$ as $P_q$ for $q \in [1, 2m+n]$. Observe that, an optimal firefighting strategy would not have $(1)$ two distinct firefighters $u_i$ and $u_j$ for $i\neq j$ on the same path $P_q$ and $(2)$ two distinct firefighters $u_i$ and $u_j$ for $i\neq j$ on the paths $P_{q_1}$ and $P_{q_{2}}$ that are incident on the same node $v_i^{(1)} \in V^{(1)}$. As for both the conditions, it is better to have one firefighter on the node $v_i^{(1)} \in V^{(1)}$ on which the paths incident. Hence, we can assume that at most one firefighter is placed on any path $P_q$ and at most one firefighter is placed on the paths incident on vertex $v_i^{(1)}$.
	Moreover, if a firefighter is placed at any node $p_i$ on a path $P_q$ between $v_i^{(1)} \in V^{(1)}$ and $v_j^{(2)} \in V^{(2)}$, then, in $2k$ time steps it can extend protections to none other than $v_i^{(1)} \in V^{(1)}$, $v_j^{(2)} \in V^{(2)}$, and the intermediate nodes on the path $P_q$. It can be said that any node on a path $P_q$ has exactly one vertex $v_i^{(1)} \in V^{(1)}$ within in a diameter of $2k$. Thus, placing a firefighter on any path $P_q$ is equivalent to placing a firefighter on the vertex $v_i^{(1)}$ on which the path $P_q$ is incident. Therefore, in the firefighting strategy $S$, if $u_i$ is a node on a path $P_q$ which is between $v_x^{(1)} \in V^{(1)}$ and $v_y^{(2)} \in V^{(2)}$, we push back the firefighter to $v_x^{(1)} \in V^{(1)}$. As the fire reaches $V^{(1)}$ in $2k$ time steps, there can be at most $k$ nodes $v_i^{1} \in V^{(1)}$ on which the firefighters are placed at the allowed time steps. Therefore, there is at least one vertex $v_i^{(1)} \in V^{(1)}$ that is burnt. And thus, if the firefighters are not placed on the vertices $v_i^{(1)} \in V^{(1)}$ that form a dominating set in $G$, then there is at least one path to a node $v_j^{(2)} \in V^{(2)}$ at which the firefighters cannot extend protection.
\end{proof}\begin{proof}
	Let $(G, k)$ be an instance of $k$-\textsc{Dominating Set} problem. We construct a graph $G'$ as follows. Add $2$ copies $V^{(1)}$ and $V^{(2)}$ of the nodes $V(G)$ in $G'$, i.e. for each node $v_i \in V(G)$ add nodes $v_i^{(1)}$ and $v_i^{(2)}$. Add a vertex $s$, the vertex from where the fire breaks out. For each vertex $v_i^{(1)} \in V^{(1)}$, add a path of length $k$ from $s$ (i.e. a path from $s$ to $v_i^{(1)}$ would be like $(s, u_{i_1}, \dots, u_{i_{k-1}}, v_i^{(1)})$). Similarly, for each edge $(v_i, v_j) \in E(G)$, add a path of length $k$ from $v_i^{(1)}$ to $v_j^{(2)}$, and a path of length $2k$ from $v_i^{(2)}$ to  $v_j^{(1)}$ in $G'$. Also, for each $v_i \in V(G)$ add a path of length $2k$ from $v_i^{(1)}$ to $v_i^{(2)}$. Let $C = V^{(2)}$ be the critical set and $k' = k$.

	We claim that, \SACS{} on $(G', s, k', C=V^{(2)})$ is a \YES{}-instance if and only if $k$-\textsc{Dominating Set} is a \YES{} instance on $(G,k)$.

	Suppose that, $G$ has a $k$-dominating set $K$. Then the strategy that saves the critical set $C$ is the one that protects the vertices $v_i^{(1)} \in V^{(1)}$ corresponding to the vertices $v_i$ in the dominating set $K$. The nodes $v_i \in K$ being the dominating set, the corresponding nodes $v_i^{(1)} \in V^{(1)}$ dominate all the nodes $v_j^{(2)} \in V^{(2)}$. And thus, the firefighters on the nodes $v_i^{(1)} \in V^{(1)}$ corresponding to the dominating set $K$, eventually extend firefighters to all the nodes in $V^{(2)}$ before fire reaches to any node $v_j^{(2)} \in V^{(2)}$. Also, protecting the vertices $v_i^{(1)} \in V^{(1)}$ corresponding to $v_i \in K$ is a valid firefighting strategy as per the Definition \ref{def:ff-strategy}.

	Suppose that $\mathfrak{h}: [k] \rightarrow V(G')$ is an optimal  firefighting strategy for $(G', s, k', C=V^{(2)})$. Let $S = \{u_1, \dots, u_k\}$ be the set of vertices which are protected by the firefighting strategy, i.e. $\mathfrak{h}(i) = u_i$ for $i \in [k]$. Note that, as per the definition of the firefighting game, $i^{th}$ firefighter is placed at time step $2i-1$. Lets denote the paths from the nodes in $V^{(1)}$ to the nodes in $V^{(2)}$ as $P_q$ for $q \in [1, 2m+n]$. Observe that, an optimal firefighting strategy would not have $(1)$ two distinct firefighters $u_i$ and $u_j$ for $i\neq j$ on the same path $P_q$ and $(2)$ two distinct firefighters $u_i$ and $u_j$ for $i\neq j$ on the paths $P_{q_1}$ and $P_{q_{2}}$ that are incident on the same node $v_i^{(1)} \in V^{(1)}$. As for both the conditions, it is better to have one firefighter on the node $v_i^{(1)} \in V^{(1)}$ on which the paths incident. Hence, we can assume that at most one firefighter is placed on any path $P_q$ and at most one firefighter is placed on the paths incident on vertex $v_i^{(1)}$.
	Moreover, if a firefighter is placed at any node $p_i$ on a path $P_q$ between $v_i^{(1)} \in V^{(1)}$ and $v_j^{(2)} \in V^{(2)}$, then, in $2k$ time steps it can extend protections to none other than $v_i^{(1)} \in V^{(1)}$, $v_j^{(2)} \in V^{(2)}$, and the intermediate nodes on the path $P_q$. It can be said that any node on a path $P_q$ has exactly one vertex $v_i^{(1)} \in V^{(1)}$ within in a diameter of $2k$. Thus, placing a firefighter on any path $P_q$ is equivalent to placing a firefighter on the vertex $v_i^{(1)}$ on which the path $P_q$ is incident. Therefore, in the firefighting strategy $S$, if $u_i$ is a node on a path $P_q$ which is between $v_x^{(1)} \in V^{(1)}$ and $v_y^{(2)} \in V^{(2)}$, we push back the firefighter to $v_x^{(1)} \in V^{(1)}$. As the fire reaches $V^{(1)}$ in $2k$ time steps, there can be at most $k$ nodes $v_i^{1} \in V^{(1)}$ on which the firefighters are placed at the allowed time steps. Therefore, there is at least one vertex $v_i^{(1)} \in V^{(1)}$ that is burnt. And thus, if the firefighters are not placed on the vertices $v_i^{(1)} \in V^{(1)}$ that form a dominating set in $G$, then there is at least one path to a node $v_j^{(2)} \in V^{(2)}$ at which the firefighters cannot extend protection.
\end{proof}


%% file: ffcs-conclusions.tex
In this work we presented the first FPT algorithm, parameterized by the number of firefighters, for a variant of the Firefighter problem where we are interested in protecting a critical set. We also presented a faster algorithms on trees. In contrast, we also show that in the spreading model protecting a critical set is W[2]-hard. Our algorithms exploit the machinery of important separators and tight separator sequences. We believe that this opens up an interesting approach for studying other variants of the Firefighter problem.